# Unraveling the Link between Periodontitis and Inflammatory Bowel Disease: Challenges and Outlook

Himanshi Tanwar[1, #], Jeba Mercy Gnanasekaran[1, #], Devon Allison[1], Ling-shiang Chuang[2], Xuesong He[3], Mario Aimetti[4], Giacomo Baima[4], Massimo Costalonga[5], Raymond K. Cross[6], Cynthia Sears[7], Saurabh Mehandru[2], Judy Cho[2], Jean-Frederic Colombel[2], Jean-Pierre Raufman[6], Vivek Thumbigere-Math[1, 8, *]

**Affiliations:**

[1] Division of Periodontology, University of Maryland School of Dentistry, Baltimore, MD, USA

[2] Division of Gastroenterology, Department of Medicine, Icahn School of Medicine at Mount Sinai, New York, NY, USA

[3] Department of Microbiology, The Forsyth Institute, Cambridge, MA, USA

[4] Department of Surgical Sciences, C.I.R. Dental School, University of Turin, Turin, Italy.

[5] Department of Diagnostic and Biological Sciences, School of Dentistry, University of Minnesota, Minneapolis, USA

[6] Division of Gastroenterology & Hepatology, University of Maryland School of Medicine, Baltimore, MD, USA

[7] Division of Infectious Diseases, Johns Hopkins University School of Medicine, Baltimore, MD, USA

[8] National Institute of Dental and Craniofacial Research, NIH, Bethesda, MD, USA

[*] **To whom correspondence should be addressed:** vthumbigere@umaryland.edu

[#] These authors contributed equally to the study

**Short Title:** Periodontitis and Inflammatory Bowel Disease Interconnection

Total Words: 7,016

Figures - 3

Tables – 2

Reference – 341

# Abstract

Periodontitis and Inflammatory Bowel Disease (IBD) are chronic inflammatory conditions, characterized by microbial dysbiosis and hyper-immunoinflammatory responses. Growing evidence suggest an interconnection between periodontitis and IBD, implying a shift from the traditional concept of independent diseases to a complex, reciprocal cycle. This review outlines the evidence supporting an "Oral-Gut" axis, marked by a higher prevalence of periodontitis in IBD patients and vice versa. The specific mechanisms linking periodontitis and IBD remain to be fully elucidated, but emerging evidence points to the ectopic colonization of the gut by oral bacteria, which promote intestinal inflammation by activating host immune responses. This review presents an in-depth examination of the interconnection between periodontitis and IBD, highlighting the shared microbiological and immunological pathways, and proposing a "multi-hit" hypothesis in the pathogenesis of periodontitis-mediated intestinal inflammation. Furthermore, the review underscores the critical need for a collaborative approach between dentists and gastroenterologists to provide holistic oral-systemic healthcare.

**Introduction**

Oral pathologies can profoundly impact general health; however, oral health and general health are often incorrectly perceived as separate entities[1,2]. Over the past four decades, several studies have highlighted the connection between the oral cavity and the rest of the body, linking more than 50 systemic conditions, including inflammatory bowel diseases (IBD), to gingival and periodontal inflammation[1-7]. A bidirectional relationship exists between periodontitis and IBD, where chronic inflammation in either the oral cavity or intestinal tissues can influence each other. Microbial and inflammatory changes originating in the oral cavity can potentially initiate or exacerbate IBD. The growing body of literature associating periodontitis with gastrointestinal (GI) disorders such as IBD and colorectal cancer has seen a remarkable surge in recent years, with over 100 relevant articles published since 2020. This review aims to augment current knowledge by providing an in-depth analysis of the underlying mechanisms contributing to the bidirectional relationship between periodontitis and IBD. We utilize the "oral-gut axis" as a framework to investigate the reciprocal relationship between the periodontium and the GI tract. This narrative review encapsulates historical studies on oral manifestations in IBD, and explores recent advances, current knowledge gaps, and potential future directions. Furthermore, the review underscores the critical need for a collaborative approach between dentists, gastroenterologists, immunologists, and infectious disease experts/microbiologists to provide holistic oral-systemic healthcare.

**Periodontitis**

Periodontitis is a chronic inflammatory disease that affects the periodontium, encompassing tooth-supporting structures such as the gingiva, alveolar bone, and periodontal ligament[7-10]. It is ranked as the 11th most prevalent condition worldwide, affecting approximately

50% of adults over 30 years of age and about 70.1% of adults over 65 years of age[11-15]. The hallmark symptoms of periodontitis are gingival inflammation, gingival recession, halitosis, deep periodontal pockets, clinical attachment loss, and tooth mobility[7-10]. If left untreated, periodontitis results in tooth loss, and is the leading cause of tooth loss in adults worldwide[10,11,16,17].

Periodontitis is primarily initiated by the accumulation of bacterial plaque on tooth surfaces or within the gingival sulcus, which subsequently triggers an inflammatory response in the periodontium[9,18]. The host's hyper-immunoinflammatory response, in turn, leads to the destruction of tooth-supporting structures[19-21]. The bacterial plaque/calculus associated with the initiation of periodontitis specifically contains bacteria, such as *Porphyromonas gingivalis*, *Tannerella forsythia*, *Treponema denticola*, *Aggregatibacter actinomycetemcomitans*, *Prevotella intermedia*, *Fusobacterium nucleatum*, and *Campylobacter rectus*[18-20,22-26]. Periodontitis is a polymicrobial disease primarily driven by bacteria; however, fungi, protozoa, and viruses might also contribute to its pathogenesis[9,27,28].

Several risk factors have been associated with periodontitis, including poor oral hygiene, smoking, diabetes, advanced age, genetics, and certain medications[7,29-31]. The prognosis depends on disease severity, patient compliance with treatment, and control of risk factors[32]. Through dissemination of oral pathogens and promotion of low-grade systemic inflammation, periodontitis may contribute to other systemic diseases such as IBD, cardiovascular disease, diabetes, and pre-term birth[33-38]. Periodontitis management focuses on infection control, reduction of inflammation, and preservation of tooth-supporting structures. Treatment strategy includes a sequential approach - (a) Control of local and systemic risk factors such as smoking and uncontrolled diabetes mellitus;

(b) Non-surgical therapy: scaling and root planing to remove bacterial plaque and calculus; (c) Medications: antimicrobial therapies, including local or systemic antibiotics, antiseptics, and host-modulating agents to manage bacterial infection; and (d) Surgical treatments: flap surgery, bone or soft tissue grafts, and guided tissue regeneration to manage advanced cases of periodontitis[39].

**Inflammatory Bowel Disease**

IBD encompasses a group of chronic, idiopathic, and recurring inflammatory disorders affecting the GI tract. The two primary forms of IBD are Crohn's disease (CD) and ulcerative colitis (UC)[40-42]. The clinical manifestations of IBD vary depending on the location and severity of inflammation within the GI tract. CD is characterized by patchy inflammation that can affect the full thickness of the bowel wall throughout the small and large intestines[42-45]. There may be areas of healthy tissue between inflamed sections (skip lesions). In contrast, UC presents with continuous, uniform inflammation limited to the mucosal layer, extending from the rectum throughout the colon[42-46]. Common symptoms include abdominal pain, cramping, persistent diarrhea, nausea, vomiting, weight loss, fatigue, and fever[40-42,47]. UC more likely causes bloody diarrhea, urgency to defecate, and tenesmus[46-49]. Approximately 25-40% of UC and CD patients may exhibit extraintestinal manifestations such as arthritis, erythema nodosum, pyoderma gangrenosum, oral lesions (discussed in detail in later sections), uveitis, hepatic steatosis, primary sclerosing cholangitis, and metabolic bone disease[50-67]. Both UC and CD patients are at a higher risk of developing colorectal cancer, although the risk is greater in UC than CD[68,69]. IBD patients experience a diminished quality of life and have a lower life expectancy compared to the general population[70,71].

Between 1990 and 2017, the global incidence of IBD surged by 31%, affecting around 6.8 million people worldwide and becoming a major public health concern[72-75]. The incidence of IBD varies, with highest rates observed in North America, Western Europe, and Australia[74,75]. The prevalence of IBD in the United States is estimated to be over 1.6 million, with 780,000 diagnosed with CD and 907,000 with UC[73,76,77].

The etiology of IBD remains largely unclear, but it is believed to result from a complex interplay between genetic, environmental, microbial factors, and the host immune responses[78-83]. Over 200 genetic loci related to immune regulation, intestinal barrier function, and other biological pathways have been associated with IBD, suggesting a strong genetic influence[78,84-89]. Notable genetic loci include *NOD2* (one of the first genetic loci identified as a risk factor for CD), *IL23R*, *ATG16L1*, *IRGM*, *PTGER4*, *TNFSF15*, and *IRF8*[78,84-89]. Previously, our group and others have identified several single nucleotide polymorphisms (SNPs) located 1.7-60kb downstream of *IRF8* as risk factors for UC and CD[87,88,90-100] **(Figure 1)**. *IRF8* SNPs are significantly associated with CD in Ashkenazi Jews, who have a four- to seven-fold-higher incidence of CD than non-Ashkenazi European Jews[96,101]. IRF8 is expressed in the intestinal epithelial cells (IECs) in a gradient pattern, with highest expression in the differentiated cell zones and low levels in the proliferating and stem cell zones[102,103]. IRF8 functions as a suppressor of colonic inflammation. IRF8 deficiency in IECs is linked to colitis-related colon tumorigenesis in both humans and mice[102,104,105]. In mice, IRF8 deficiency promotes colitis mediated by Th17 and Tfh cells[104,105] and impairs gastric innate immunity against infection[106]. Dysregulation of IRF/IFN-I signaling is extensively involved in IBD pathogenesis and targeting this pathway represents a promising intervention strategy for IBD[107]. Additionally, we have shown that IRF8 is also critical for periodontal homeostasis[108-110].

Collectively, these findings highlight the importance of IRF8 in intestinal health and establish the critical value of *Irf8*-deficient mice in providing novel insights into human GI diseases.

Some of the identified risk factors for IBD include: (a) Family history: individuals with a first-degree relative with IBD face a higher risk of developing the condition[111,112]; (b) Age: IBD is more commonly diagnosed in individuals under the age of 30 years, although it can occur at any age. Historically believed to be a disease of children and young adults (20-30 years), it is now recognized that 10-15% of patients develop IBD after the age of 60 years (older adults)[113-117]. The cause of older-onset IBD is unclear, but it is thought to be less associated with genetics and more influenced by age-related changes in the immune system and gut microbiome[113,118-121]; (c) Ethnicity: IBD is more common in Caucasians and people of Ashkenazi Jewish descent[96,101,111,122,123]; and (d) Smoking: cigarette smoking increases the risk of CD and may exacerbate its progression, while it appears to have a protective effect against UC[124,125].

The course of IBD varies, with some experiencing periods of remission while others endure frequent relapses. IBD may lead to complications such as strictures, fistulas, or colorectal cancer[40-47]. Disease severity and treatment response significantly influence prognosis. Currently, there is no cure for IBD, and the primary management focuses on resolving inflammation, preventing flare-ups and other complications, and improving quality of life[126-128]. Treatment options include - (a) Medications: anti-inflammatory drugs, immunosuppressants, biologic therapies, and small molecules[129-131]; (b) Nutrition: dietary modification and nutritional supplementation may help manage symptoms and promote healing[132-134]; and (c) Surgery: in cases of severe inflammation, strictures, or fistulas, surgery may be necessary to remove the affected portion of the intestine[135-137].

**Oral Manifestations of IBD**

Extraintestinal manifestations of IBD, which can affect nearly all organ systems, are common in both UC and CD[50-55]. Approximately 10-30% of IBD patients exhibit oral manifestations of the disease, which can precede, coincide with, or follow the onset of GI symptoms[56-58,63,64,67]. Commonly observed oral manifestations include aphthous ulcers, cobblestoned oral mucosa, pyostomatitis vegetans, gingivitis, periodontitis, angular cheilitis, and oral lichen planus[37,56-67,138,139]. Additionally, some individuals may present with halitosis, atrophic glossitis, burning mouth syndrome, and xerostomia[56-67]. Some of these lesions are more common in CD versus UC, and more frequently noted in children versus adults[140].

The etiology of oral manifestations in IBD is complex and not fully understood; however, it is believed to involve a combination of genetic predisposition, immune dysregulation, and alterations in the oral microbiome[56-67]. Several studies underscore the role of cytokine activity in both the GI and oral regions. Elevated levels of IL-6, IL-8, IL-1β, TNF-α, and MCP-1 have been noted in saliva and gingival tissues of patients with active IBD versus nonactive IBD or healthy controls[141-144]. Higher levels of chemokines CXCL-8, -9, and -10 have been identified in the buccal mucosal of children with CD compared to healthy children and adult CD patients[145]. Mutations in IL-10RA and IL-10RB, which disrupt the IL-10/STAT3 cascade, have been associated with pediatric IBD[146,147]. Patients deficient in IL-10 and IL-10R may exhibit recurrent aphthous ulcers due to subtle immunological abnormalities, including decreased CD4+/CD8+ T-cell ratios and dysregulated serum immunoglobulin levels[148,149]. Schmidt et al. identified higher concentrations of activated matrix metalloproteinase-8 (MMP-8) in the gingival crevicular fluid (GCF) of IBD patients compared to healthy controls[150]. Figueredo et al. reported decreased IL-4 levels and elevated serum IL-18 levels in UC patients, with the latter positively correlated with IL-1β in the GCF, which likely contributed to periodontitis development[151]. Likewise, Enver et al. reported a significant elevation in IL-1β concentrations within GCF, alongside increased TNF-α and reduced IL-10 levels in the saliva of UC patients diagnosed with periodontitis[152]. Low levels of IL-10 are unable to inhibit the production of pro-inflammatory cytokines, such as IL-6, TNF-α, IFN-γ, and IL-17, resulting in higher IL-17/IL-10 ratio that may serve as a key indicator of disease severity and progression due to its association with intestinal and extra-intestinal features[153,154].

Conversely, IBD may immunologically trigger oral manifestations, partially due to the recognition of common epitopes throughout the body. The extraintestinal manifestations of IBD could stem from a broad adaptive immune response triggered by local intestinal dysbiosis, leading to recognition of these epithelial epitopes in other organs and the oral cavity, ultimately causing extraintestinal pathologies[54,155-157]. A molecular similarity between gut microbiota antigens and nonmicrobial epitopes present on cells in the oral cavity could potentially lead to immune cross-reactivity[158,159]. Furthermore, in IBD, local immune responses to gut dysbiosis may initiate systemic T cell-mediated responses and cytokine production[160,161], which could lead to induction of oral lesions. In response to antigenic stimulation, T cells migrate to the oral mucosa, where CD8+ T lymphocytes, accompanied by infiltrating macrophages and neutrophils, may cause epithelial damage and ulceration commonly observed in pyostomatitis vegetans[162]. Overexpression of pro-inflammatory cytokines such as IL-6, IL-8, and TNF-α in pyostomatitis vegetans can lead to the recruitment of inflammatory cells to UC lesions, thus synergistically contributing to the pro-inflammatory pathogenesis of pyostomatitis vegetans and IBD[162]. Collectively, these studies suggest that immune system dysregulation may be an important link bridging the oral cavity and the intestines.

Several studies have suggested that oral microbial dysbiosis, and secondary inflammatory responses within the gut, might potentially contribute to oral manifestations in IBD[54,63,163,164]. Remarkably, the periodontal inflammation in IBD does not appear to correlate with plaque accumulation, giving rise to the notion that systemic inflammation associated with IBD could induce alterations in the oral microbiome, consequently intensifying oral inflammation[165,166]. An overabundance of specific oral bacteria, including *Streptococcus*, *Prevotella*, *Veillonella*, and *Haemophilus* in the oral cavity is linked to inflammatory responses triggered by reduced salivary lysozyme and increased IL-1β levels, which may be associated with gut microbial dysbiosis[141]. These changes are connected to heightened oxidative stress and virulence (e.g., metabolism of terpenoids, polyketides, carbohydrates and lipids, and biosynthesis of secondary metabolites), enzyme activity (e.g., protein kinases), bacterial aggression (e.g., bacterial toxins and invasion of epithelial cells), and apoptosis in the oral region, implying a connection between oral microbial dysbiosis and IBD[167]. Altered salivary lysozyme levels could be related to gut imbalance and subsequent periodontitis development[168]. Furthermore, intestinal dysbiosis has been linked to chronic inflammatory state and activation of gut-associated lymphoid tissue (GALT)[169,170], which could potentially lead to extraintestinal pathologies[171]. The proven effectiveness of adjuvant probiotic therapy in treating oral ulcers supports the hypothesis that oral ulcers in IBD might result from a combination of intestinal dysbiosis and other factors, such as oral mucosa microtraumas[172].

Collectively, these findings indicate a possible correlation between oral and gut microbial dysbiosis and the occurrence of oral manifestations in IBD.

**Historical Perspective of Oral-Gut Interconnection**

Over the years, numerous case reports have contributed to our understanding of oral and gut interconnection. The first report of oral involvement in IBD, by Dudeney and Todd in 1969, described a CD patient who developed granulomatous lesions in the left buccal mucosa 16 years after the initial diagnosis[173]. In 1972, Bottomely et al. documented a case of teenage girl with CD who exhibited gingival hyperplasia across maxillary anterior teeth, with 5-6 mm "pseudopockets" indicative of a severe hyperplastic phenotype that worsened periodically[174]. In 1972, Croft et al. retrospectively analyzed 332 CD patients and found that 6.1% had developed oral ulcers at some point during their illness[175]. In 1975, a systematic study by Asquith et al. involving 100 CD patients, 100 UC patients, and 100 healthy controls matched for age, sex, and denture status reported that 9% of CD and 2% of UC patients developed oral lesions with macroscopic and histological features similar to those in the GI tract[163]. Additionally, they noted salivary IgA production was reduced in CD patients with active bowel disease[163]. In 1982, Lamster et al. analyzed polymorphonuclear leukocytes from 30 patients with active or inactive IBD and reported that those with active IBD had higher levels of circulating immune complex activity and their peripheral neutrophils exhibited greater metabolic activity compared to inactive IBD or healthy controls[176]. Additionally, a higher prevalence of oral pathologies was observed among subjects with active IBD. In 1986, Van Dyke et al. evaluated 20 IBD patients with and without periodontal disease and found that the periodontal flora of IBD patients predominantly consisted of small, motile, gram-negative rods, closely related to the genus *Wolinella*[177]. All 10 IBD subjects with periodontal disease exhibited serum-mediated defect in neutrophil chemotaxis, while neutrophil phagocytosis was normal. Additionally, PGE2 levels in GCF from IBD patients with periodontal disease were four-fold higher than in adult periodontitis patients without IBD[177].

Building upon these initial case reports, cross-sectional studies were designed to explore the positive correlation between IBD and periodontitis-afflicted individuals. In 1991, Flemmig et al. examined the periodontal status of 107 IBD patients and identified a 12% higher prevalence of periodontal disease but a 0.6 mm lower clinical attachment loss in IBD patients when compared to the general US adult population[178]. In 2006, Grössner-Schreiber et al. examined 62 IBD patients and 59 healthy controls and found that IBD patients exhibited a significantly higher incidence of caries, but no distinct differences in periodontal findings[179]. In contrast, Brito et al. in 2008 examined 99 CD, 80 UC, and 74 healthy controls and found a significantly higher prevalence of periodontitis in UC (90%) and CD (82%) patients when compared to healthy controls (68%)[138]. Altogether, these studies provide some of the earliest evidence for the oral-gut axis, highlighting the potential connection between periodontitis and IBD. Over the next several years, similar studies have been published expanding our knowledge about periodontitis-IBD interconnection. **Table 1** summarizes recent studies demonstrating associations between periodontitis and IBD.

Currently, limited evidence suggests that IBD treatment can improve periodontal disease outcomes, and vice versa, hinting that both diseases might share common altered immune-inflammatory mechanisms. IBD patients treated with anti-TNF-α biologics experienced rapid healing of apical periodontitis compared to controls[180]. Corticosteroids, which are often used in IBD management, seem to confer a protective effect against periodontitis[181]. In UC patients responsive to biologics treatment, salivary levels of IgA and MPO significantly increased, suggesting that successful UC therapy may also boost oral defense mechanisms[182]. Furthermore, biologics used to treat other chronic inflammatory diseases appear to hinder periodontitis progression and enhance the healing response to periodontal treatment[183]. Conversely, periodontitis treatment with mesenchymal stem cell-derived exosomes attenuated experimental colitis in murine models[184]. Taken together, these findings indicate that the immune-modulating effect of certain drugs might not only improve IBD outcomes, but also reduce inflammation within periodontal tissues. Moreover, periodontitis treatment could have positive effects on IBD outcomes. Nonetheless, it's crucial to note that immunomodulators commonly used in IBD management, such as corticosteroids, aminosalicylates, methotrexate, anti-TNF-α, and other biologics, could increase the risk of opportunistic infections in the oral cavity[65].

**A Multi-hit Model of Periodontitis-Mediated IBD Pathogenesis**

Every day, the average adult generates and ingests ~1.5 L of saliva, which contains ~$1.5 \times 10^{12}$ oral bacteria[185-189]. In healthy individuals, several defense mechanisms prevent ingested oral bacteria from colonizing the gut and promoting IBD. These include: (a) physical barrier provided by the mucus layer along with IECs[190], (b) bacterial clearance by mucus[191], (c) competition between indigenous gut microbes and exogenous microbes invading from the mouth[192,193], and (d) gastric acid and pepsin destroy opportunistic pathogens that enter the stomach[194,195]. However, when these defense mechanisms are breached by genetic abnormalities, systemic disorders, gut dysbiosis, medications, and other factors, oral microbes can disseminate to the intestine, potentially disrupting intestinal homeostasis and abnormally activating the intestinal immune system, thereby influencing IBD development[3,193,196-203].

The mucus layer rich in antimicrobial peptides and secretory immunoglobulins prevents direct contact between pathogens and the epithelium and creates a hostile environment for potential invaders[204,205]. Conditions such as cystic fibrosis, stress, infection, an unhealthy diet, and aging can disrupt the mucus layer, facilitating bacterial penetration[120,206-208]. Beneath the mucus, the IEC barrier, fortified by occludins, claudins, and other junction proteins, provides another layer of protection[209]. Disruption of the IEC barrier can result in a "leaky" gut, permitting microorganisms and intestinal contents to infiltrate the mucosal barrier and initiate inflammation[210]. Further, the resident commensal microbiota resist exogenous pathogen colonization through nutrient competition and niche occupation[192,211,212]. These microbes ferment dietary fibers, producing short-chain fatty acids (SCFAs) that serve as an energy source for colonocytes and exhibit anti-inflammatory properties[213]. However, a shift in microbial equilibrium, or dysbiosis, frequently induced by triggers such as antibiotic overuse or dietary changes, can exacerbate IBD risk[214]. Chemical barriers, including gastric acid and pepsin in the stomach, help thwart opportunistic pathogens[194,195]. However, hypochlorhydria or prolonged use of proton pump inhibitors may predispose individuals to bacterial overgrowth and infections like *Clostridium difficile*[215]. Collectively, the compromised mucus layer and epithelial barriers, coupled with a dysbiotic state of gut microbiota and heightened immune responses, can increase the risk for IBD. Based on these findings, we propose a multi-hit model to illustrate how periodontitis and related oral bacteria may enhance the susceptibility to IBD **(Figure 2)**.

**Stage 1: Initiation of Oral Dysbiosis**

In a host exhibiting homeostatic oral and gut compartments, oral bacteria neither proliferate nor translocate to the gut, thus maintaining a state of health. The onset of periodontitis causes oral dysbiosis, leading to the expansion of virulent oral bacteria[216]. There is an enrichment of gram-negative bacteria that can thrive within anaerobic conditions, such as those found in the gut. Documentation of genetically identical strains of *F. nucleatum* and *C. concisus* in both the oral cavity and the intestines of IBD and colorectal cancer patients indicates microbial adaptations that promote oral-to-gut transmission[217,218].

**Stage 2: Oral Bacteria Translocation and Intestinal Colonization**

The next stage involves oral bacteria translocation to the intestine, either via ingestion or through the circulatory system (bacteremia). Normally, more than 99% of swallowed oral microorganisms are inactivated as they pass through the stomach[189,219]. However, several conditions such as gastric ulcers, gastroesophageal reflux disease, genetic disorders, an unhealthy diet, aging, stress, and the use of tobacco, alcohol, antibiotics, non-steroidal anti-inflammatory drugs, and/or proton-pump inhibitors may disrupt the intestinal barrier function and perturb the gut microbiota ecosystem, thus disrupting colonization resistance and facilitating the establishment of oral bacteria in the gut[120,206-208,214,215,220-222]. A 10-fold increase in ileal *Fusobacteriaceae* was noted following antibiotic usage[223]. Similarly, increased intestinal colonization of oral bacteria was noted in individuals using proton pump inhibitors[224,225]. Additionally, periodontitis can lead to bacteremia, with *P. gingivalis, F. nucleatum, T. denticola,* and *P. intermedia* capable of evading immune surveillance and proliferating within the immune cells[9,226], and potentially exacerbating IBD development.

**Stage 3: Induction of Intestinal Inflammation**

In homeostasis, gut-resident microbes promote development of bacterium-specific tolerogenic responses against commensals, that do not elicit intestinal inflammation[227,228]. However, when certain exogenous oral pathogens colonize the gut, they can opportunistically elicit pathogenic immune responses. After epithelial disruption, ingested oral bacteria can interact with gut-associated immune cells stimulating production of pro-inflammatory cytokines such as IL-17, IL-1β, and IFN-γ[200,229]. Furthermore, oral pathogen-reactive Th17 cells that arise *de novo* in the oral cavity can migrate to the inflamed gut, where they are activated by ectopically colonized oral pathogens, and subsequently contribute to gut inflammation[200]. Thus, periodontitis can exacerbate gut inflammation by supplying the gut with both colitogenic oral pathogens and pathogenic T cells[200].

In summary, having oral dysbiosis or gut barrier disruption alone is insufficient to render the host susceptible to IBD. However, the simultaneous presence of oral dysbiosis (providing an adequate supply of oral pathogens) and impaired gut resistance (disrupted barrier function against oral pathogens) creates necessary conditions for successful gut colonization by oral pathogens, which then promotes intestinal inflammation by activating innate and adaptive immune responses.

**Oral and Gut Microbiome**

The oral microbiota ranks as the second-largest microbial community after the gut microbiota. It harbors more than 700 species and 1,300 bacterial strains, along with a diverse array of ultrasmall Candidate Phyla Radiation bacteria, fungi, amoebae, flagellates, archaea, and viruses[24,27,186,230-237]. In health, the oral microbiota is remarkably stable and dominated by commensal bacteria such as *Streptococcus*, *Actinomyces*, *Haemophilus*, and *Neisseria*, which maintain a dynamic equilibrium with the host, ensuring symbiosis and facilitating normal immune and metabolic functions within the oral cavity[9,18,20,232,238]. However, when this balance is disrupted, it can lead to dysbiosis, resulting in periodontitis[9,18,20,25,232,238]. Oral dysbiosis is characterized by the overgrowth of anaerobic bacteria such as *P. gingivalis*, *T. forsythia*, *T. denticola* (collectively referred to as the "red complex"), and other species within the phyla *Firmicutes*, *Proteobacteria*, *Spirochaetes*, and *Bacteroidetes*[9,18,20,25,232,238]. These dysbiotic bacteria produce virulence factors that promote tissue destruction, impair host immune defenses, and disrupt the oral microbial community's stability[18]. While the dysbiotic microbiota enriched in virulence factors triggers periodontitis, it is the host's hyperactive immunoinflammatory responses to the altered microflora that cause major destruction of tooth-supporting structures[19,21].

*P. gingivalis* is a keystone periodontal pathogen, which adheres to mucus membranes, periodontal pocket epithelium, and other bacterial surfaces through adhesion factors such as fimbriae, hemagglutinins, proteases, and adhesins[239-241]. *P. gingivalis* produces a range of virulence factors, including endotoxins (e.g., lipopolysaccharide (LPS)), proteases (e.g., gingipains), outer membrane vesicles, acid and alkaline phosphatases, and organic acids, that contribute to the degradation of host proteins and evasion of host immune defenses[240-243]. Consequently, this leads to clinical manifestations such as edema, neutrophil infiltration, and hemorrhage. In addition to *P. gingivalis*, other periodontal pathogens, such as *A. actinomycetemcomitans*, *F. nucleatum, T. forsythia*, and *P. intermedia*, play crucial roles in the pathogenesis of periodontitis[23,244-248]. *A. actinomycetemcomitans* is a facultative anaerobe and its key virulence factors include leukotoxin (LtxA), which selectively targets immune cells, cytolethal distending toxin (Cdt) that induces cell cycle arrest and apoptosis in multiple cell types, and adhesins that promote bacterial attachment to host tissues and other oral microbiota[246-250]. Additionally, its LPS triggers inflammation, and the extracellular matrix-degrading enzymes lead to tissue destruction, together orchestrating the complex pathophysiological processes underlying periodontitis. *F. nucleatum* is an anaerobic bacterium known to promote biofilm formation and support the growth of other periodontal pathogens[244,245,251]. It produces various virulence factors, including adhesins, hemagglutinins, and proteases, that contribute to tissue destruction and evasion of host defenses[252,253]. *T. forsythia*, another important periodontal pathogen, produces virulence factors such as proteases (e.g., karilysin and mirolase) and surface structures like the BspA protein, which facilitate tissue invasion and immune evasion[254,255]. The combined actions of multiple periodontal pathogens and their virulence factors contribute to the molecular mechanisms underlying periodontitis[240-255].

As two extremities of the digestive tract, the oral cavity and the intestine exhibit shared microbial pathogenesis mechanisms. Imbalances in the gut microbial community can precipitate alterations in bacterial metabolites and the release of numerous virulence factors, leading to host tissue damage and inflammatory responses[192,256-258]. In a healthy state, the gut microbiota is primarily composed of *Firmicutes*, *Bacteroidetes*, *Proteobacteria*, and *Actinobacteria*, with smaller populations of *Fusobacteria* and *Cyanobacteria* contributing to the overall diversity[256,259-261]. Additionally, archaea, fungi, and viruses colonize the human GI tract. This intricate microbial community maintains gut homeostasis by executing essential functions such as nutrient metabolism, immune modulation, and defense against pathogens[256,259-263]. The host tissues offer a nutrient-rich environment, reciprocated by the gut microbiota through the production of SCFAs and essential vitamins, fostering a mutually beneficial symbiosis[261,263]. However, in IBD, the gut microbiota undergoes ecological dysbiosis, characterized by reduced microbial diversity, a decrease in beneficial commensal bacteria (e.g., Firmicutes), and an increase in potentially pathogenic bacteria (e.g., Proteobacteria)[256,259-261]. This dysbiosis leads to reduced production of beneficial metabolites such as SCFAs and an increase in pro-inflammatory molecules, which can further exacerbate the intestinal inflammation and compromise the intestinal barrier function[256,259-261]. As a result, the delicate balance between the host and the gut microbiota is disrupted, leading to a vicious cycle of inflammation and tissue damage that characterizes IBD.

Clinical and animal studies have demonstrated an emerging microbial signature in IBD, characterized by the expansion of oral-associated bacteria. **Table 2** summarizes various clinical studies demonstrating evidence of oral bacterial species enriched in the intestinal tissues of patients with IBD and colorectal cancer (another disease implicated in the oral-gut axis). In an early study, Van Dyke et al (1986) investigated differences in the oral microbiome among IBD, periodontitis patients, and healthy controls[177]. In IBD patients, they observed a predominant presence of gram-negative rod-shaped bacteria, closely related to the genus *Wolinella*. Interestingly, periodontitis-free IBD patients displayed a decreased bacterial load within the gingival sulcus compared to their counterparts with periodontitis. These findings suggest distinct variations in the oral microbiome of IBD patients compared to periodontitis patients and healthy controls. Meurman et al. (1994) found higher salivary yeast, *Lactobacilli*, and *S. mutans* counts in individuals with active CD compared to inactive CD[264]. Stein et al. (2010) detected higher frequencies of *C. rectus* (94.6%), *A. actinomycetemcomitans* (76.9%), *P. gingivalis* (62.6%), *P. intermedia* (79.6%), and *T. forsythia* (64.6%) in subgingival plaque from 147 CD patients stratified for *CARD15*-gene mutations[265].

Notably, this study did not include a control group for comparative analysis. Man et al. (2010) and Strauss et al. (2011) identified a higher abundance of *C. concisus* and *F. nucleatum* in feces and intestinal biopsies of IBD patients, suggesting that viable *C. concisus* and *F. nucleatum* can migrate to the intestines and establish themselves in the mucosal environment[266,267]. Docktor et al. (2012) investigated the oral microbiome of children and young adults with IBD by analyzing tongue and buccal mucosal brushings[268]. They found a significant decrease in *Fusobacteria* and *Firmicutes* in CD patients compared to healthy controls. In UC patients, a similar decrease in *Fusobacteria* was observed, while *Spirochaetes*, *Synergistetes*, and *Bacteroidetes* were increased compared to controls. Brito et al. (2013) investigated 45 untreated periodontitis patients with either CD or UC, and patients with untreated periodontitis as controls[38]. They found CD patients harbored significantly higher concentrations of *B. ureolyticus*, *C. gracilis*, *P. melaninogenica*, *S. aureus*, *S. anginosus*, *S. intermedius*, and *S. mutans* compared to UC and controls subjects. Said et al. (2014) analyzed saliva from 21 CD, 14 UC patients, and 24 healthy controls, and found higher abundance of *Bacteroidetes*, *Prevotella*, and *Veillonella* in CD and UC patients compared to healthy controls, while *Proteobacteria*, *Streptococcus*, and *Haemophilus* were lower[141]. *Neisseria* and *Gemella* were also lower in CD patients versus healthy controls. Kelsen et al. (2015) characterized subgingival microbiota in pediatric patients with active or non-active CD and identified 17 genera more abundant in CD patients versus healthy controls, including *Capnocytophaga*, *Rothia*, and *TM7*[269]. Both antibiotic exposure and disease state were linked to differences in bacterial community composition. Schmidt et al. (2018) examined subgingival plaque and found a significantly lower prevalence of *E. nodatum* and *E. corrodens* in 59 IBD patients compared to 59 healthy controls[150]. Xun et al. (2018) investigated oral microbial dysbiosis in saliva samples from 54 UC patients, 13 active or remissive CD patients, and 25 healthy adults. They found an enrichment of *Streptococcaceae* and *Enterobacteriaceae* in UC, *Veillonellaceae* in CD, and depletion of *Lachnospiraceae* and *Porphyromonadaceae* in UC, as well as *Neisseriaceae* and *Haemophilus* in CD patients[167]. Alpha diversity was significantly lower in UC and CD patients. Recently, Kitamoto et al. (2020) showed that *Klebsiella* spp. from the oral cavity can translocate to the gut, where they activate the inflammasome in lamina propria macrophages, exacerbating gut inflammation[200].

**Shared Immunoinflammatory Responses**

The pathogenesis of both periodontitis and IBD involves a complex interplay between host immunity and specific bacterial stimuli[270]. Here, we discuss the sequence of events involving immune cell activation by specific oral bacteria contributing to periodontitis-mediated IBD progression, elaborating on the mechanistic details of the roles of neutrophils, macrophages, dendritic cells (DCs), T cells, and B cells in the process. Oral pathogen-mediated immune responses that drive gut inflammation in IBD are concisely illustrated in **Figure 3**.

**Neutrophils**, as the first responders to bacterial invasion, play a pivotal role in the initial stages of periodontitis and IBD[9,271-275]. In periodontitis, *P. gingivalis*, through its virulence factors such as gingipains, promotes a hyperactive and sustained neutrophilic response, leading to elevated release of proteolytic enzymes and reactive oxygen species[9,271,276]. This results in collateral tissue damage and subsequent alveolar bone loss, a hallmark of periodontitis. Moreover, *P. gingivalis* can dysregulate neutrophil function, such as impairing neutrophil extracellular trap (NET) formation and phagocytosis, leading to an imbalanced host response[9,271,276]. In IBD, similar neutrophil dysfunctions are observed, where excessive and sustained neutrophilic infiltration in the intestinal mucosa contributes to chronic inflammation and tissue damage[273,274,277]. Furthermore, neutrophil dysregulation, including abnormal recruitment and impaired clearance, is also seen in IBD[273,274,277]. *P. gingivalis* may dysregulate neutrophil function, potentially exacerbating IBD pathogenesis. Systemically circulating neutrophils in individuals with periodontitis exhibit cytokine hyper-reactivity and impaired chemotaxis, which could potentially contribute to the oral–gut axis in the context of IBD[176,177,278]. S100A8, S100A9, and S100A12 are small calcium-binding proteins abundantly expressed by neutrophils in acute inflammation and have been implicated in immune regulation in both periodontitis and IBD[279-281]. Higher salivary expression of S100A12 has been found in UC patients with periodontitis[282]. The role of S100 proteins in periodontitis-IBD pathogenesis needs to be further investigated.

**Macrophages**, key players in immune response, are implicated in both periodontitis and IBD. In the context of periodontitis, *P. gingivalis* can influence macrophage polarization, favoring a pro-inflammatory (M1) phenotype over an anti-inflammatory (M2) phenotype, thus promoting inflammation and tissue damage[283]. *F. nucleatum* also stimulates M1 macrophage polarization via the AKT2 pathway and fosters Th1 and Th17 cell expansion through STAT3 signaling[284,285]. Similar macrophage polarization patterns are seen in IBD, contributing to a sustained inflammatory environment in the gut[286,287]. Specifically, *Enterobacteriaceae* such as *Klebsiella* isolated from the oral cavity trigger IL-1β secretion from macrophages, intensifying intestinal inflammation by mediating IL-1 signaling[200]. Intriguingly, *Enterobacteriaceae* isolated from the intestines do not evoke similar IL-1β secretion, suggesting a specific response to oral microbes[200].

**Dendritic cells (DCs)** are antigen-presenting cells that bridge innate and adaptive immunity[288,289]. DCs play a crucial role in maintaining oral tolerance to commensal bacteria and preventing unwanted inflammatory responses[290]. However, certain periodontal pathogens, like *P. gingivalis* and *A. actinomycetemcomitans*, can dysregulate DC function and promote inflammation[291,292]. In IBD, dysregulated DCs can initiate and perpetuate intestinal inflammation through an exaggerated T cell response[293,294]. Saliva from CD and UC patients can cause intestinal inflammation in mice lacking IL-10 through *Klebsiella*-induced DC signaling and Th1 activation[229]. In response to oral bacteria, IFN-γ+ CD4+ T cells accumulate in the intestinal mucosa of these mice, as observed in humans with CD.

**T cells** are key orchestrators of immune responses in both periodontitis and IBD, with their activation and subsequent modulation of inflammation serving as common pathogenic threads. *P. gingivalis* stimulates aberrant CD4+ T cell responses, tilting the balance towards a pro-inflammatory milieu. More specifically, *P. gingivalis* incites an increase in Th17 cell responses, characterized by distinct production of IL-17 and simultaneous modulation of the Th17/Treg ratio[295,296]. This results in an augmentation of Th17-related transcription factors and pro-inflammatory cytokines IL-17 and IL-6 via the TLR4 pathway, paralleled by downregulation of Foxp3, TGF-β, and IL-10[297]. Further, *P. gingivalis* elicits intestinal inflammation by altering the gut microbiota composition and disrupting epithelial barrier function through IL-9-producing CD4+ T cells[202]. These shifts towards pro-inflammatory responses and a skewed T cell landscape could contribute to IBD progression[296,297]. *F. nucleatum* exacerbates this inflammatory cascade by compromising the integrity of the intestinal epithelial barrier, thereby increasing permeability and promoting secretion of inflammatory cytokines such as TNF-α, IFN-γ, IL-1β, IL-6, IL-17, and in parallel inhibiting the production of the anti-inflammatory cytokine IL-10[285]. In periodontitis, oral pathogen-reactive Th17 cells arise *de novo* in the oral cavity and migrate to the inflamed gut, where they are activated by ectopically colonized oral pathogens, and subsequently contribute to gut inflammation[200]. *Klebsiella* species, prevalent in the oral cavity of CD patients, have been shown to exploit this inflammatory landscape[229]. Upon intestinal colonization, *Klebsiella* induces Th1 cell responses via the innate immune receptor TLR4, leading to IL-10 deficiency and further amplifying inflammation[229]. Interestingly, *Klebsiella* also promote the expression of interferon-inducible (IFI) genes, which may facilitate colonization and subsequent recruitment of Th1 cells, thereby promoting a Th1-skewed inflammatory response reminiscent of IBD-like colitis[229]. Numerous studies have further delineated the shared inflammatory mechanisms between periodontitis and IBD by evaluating cytokine expression in affected tissues. For example, Menegat et al.[298] and Figueredo et al.[144,151] identified heightened levels of IL-17A, IL-17F, IL-22, IL-23, IL-25, IL-33, INF-γ and IL-10 in gingival tissues of patients with IBD and chronic periodontitis. However, they found IL-1β, IL-4, IL-10, and IL-21 to be significantly elevated in patients with active IBD, suggesting an inflammation score based on IL-1β, IL-6, IL-21 and sCD40L expression, was higher in gingival tissues of active IBD patients. Furthermore, IBD patients treated with anti-TNF-α biologics had more rapid healing of apical periodontitis compared to controls[180]. Collectively, these studies underscore the intertwined relationship between periodontitis and IBD and the central role of T cell-mediated immune responses, thereby providing a clearer understanding of the common pathogenic landscape and potential targets for future therapeutics.

**B cells**, essential constituents of adaptive immunity that play a key role in the crosstalk between periodontitis and IBD, mediate their effects primarily through the production of antibodies against specific pathogens. *P. gingivalis* can stimulate B cells to produce antibodies, such as IgG and IgA, but also fosters an environment of sustained inflammation in the oral cavity[299,300]. There is emerging evidence that B cell dysregulation and inappropriate antibody responses also contribute to IBD pathogenesis[301]. Our group recently demonstrated a highly dysregulated B cell response in UC, highlighting a potential role of B cells in disease pathogenesis[302]. We found expansion of naive B cells and IgG+ plasma cells with curtailed diversity and maturation within the mucosal B cell compartment of UC patients.

**Current Knowledge Gaps and Future Directions**
**Clinical and animal studies**

Although evidence supports the notion that periodontitis and IBD are interrelated, the specific mechanisms by which oral bacteria contribute to gut inflammation and vice versa are still unclear. The directionality and causality of this relationship are obscured by confounding factors. Tobacco use, a Western diet, and lifestyle are recognized risk factors for IBD, with some genes linked to mucosal barrier function and immune regulation potentially implicated[40,82,83,303]. However, the precise role of these factors remains unclear. Much of the existing research linking periodontitis and IBD is cross-sectional, limiting our understanding of causal relationships[37,139,304-308]. There is a need for longitudinal studies that can provide insights into the temporal sequence of these conditions, and whether periodontitis precedes, follows, or develops concurrently with IBD. To date, most studies have focused on bacterial dysbiosis, and the role of fungi and viruses has been neglected. It remains ambiguous whether dysbiosis acts as a primary cause of the disease or if it is simply a by-product of changes in the immune system, metabolism, or diet. Furthermore, the microbial dissimilarity between humans and mice raises questions about the applicability of murine findings to humans. The role of various immune cells and the impact of trained immunity in both conditions remains ill-defined. The influence of age on the association between periodontitis and IBD is not fully clear. Age is a significant risk factor in the pathogenesis of both conditions, and understanding its role could potentially inform prevention and treatment strategies. Further it remains unanswered whether periodontitis treatment ameliorates IBD risk or severity, and if IBD management can reduce periodontitis occurrence. These significant knowledge gaps call for well-designed, robust, mechanistic clinical and animal studies to gain a comprehensive understanding of the interrelationships between periodontitis and IBD pathogenesis.

**Concerns regarding the dextran sulfate sodium (DSS)-induced colitis model**

While several animal models (at least 66 different kinds) have been employed to study IBD pathogenesis[309-312], we focus on the DSS-colitis model, which has been extensively used to investigate colitis pathogenesis and the potential causal link between periodontitis and IBD[200,285,296,297,309-316]. However, this model has inherent limitations that may impact the interpretation and translatability of oral and gut findings. DSS administration induces acute colonic inflammation by direct chemical injury to the intestinal epithelium[309-312]. The physical damage to the epithelial lining leads to luminal microbial ingress into the lamina propria, which stimulates innate and adaptive immune responses. This process differs from human IBD, which is a chronic condition characterized by alternating periods of inflammation and remission involving a complex interplay of genetic, environmental, and immunological factors[40,82,83,303]. While DSS is administered to cause colonic inflammation, it can have off-target effects on periodontal tissues and oral microbiota. Oz and Ebersole (2010) found that 2% DSS alone causes periodontal inflammation and significant alveolar bone loss in BALB/c mice[317,318]. Significant bone loss was detected as early as 7 weeks after initiation of DSS treatment, progressing to severe periodontitis by 18 weeks. The authors concluded that DSS impacts mucosal tissue in a more generalized and systemic manner, affecting both the intestine and the oral cavity. Similarly, Mello-Neto et al. reported that DSS-treated rats exhibited inflammatory cells extending into periodontal connective tissues, which contained significantly elevated expression of Th1/Th2-related cytokines such as IL-1α, IL-1β, IL-2, IL-6, IL-12, IL-13, GM-CSF, IFN-γ, and TNF-α[319]. They concluded that DSS should be used with caution as it can lead to more widespread and indiscriminate lesions, and the increased pro-inflammatory cytokines in the gingival tissues caused by DSS alone might create an environment conducive to future alveolar bone loss. In another study, Rautava et al. showed that C57BL/6 male mice treated with 2% DSS for 1 week exhibited altered oral and gut microbiome composition, with a decrease in oral levels of *Spirochetes*, *Betaproteobacteria*, and *Lactobacillus*[320]. The salivary microbiota exhibited the most significant change when compared to the microbiota of the tongue and buccal mucosa. However, they noted no visible oral inflammation, and it is unclear how much of this is attributed to gender imbalance in the study. Additionally, Metzger et al.[321] and Hamdani et al.[322] demonstrated that DSS treatment suppresses bone formation and increases bone resorption, resulting in reduced bone mass and altered bone microarchitecture. In DSS-treated mice, elevated TNF-α, IL-6, RANKL, OPG, and sclerostin corresponded with higher osteoclast surfaces and lower rates of bone formation. Collectively, these findings suggest that DSS promotes periodontal inflammation, alveolar bone loss, and altered oral microbiota.

Based on these findings, the question arises whether DSS is a confounder in studying periodontal-IBD interconnection. While seminal studies in DSS-treated mice showed that bacteria and immune cells from the oral cavity migrate to the gut and exacerbate colonic inflammation[200], the extent to which this inter-relationship is confounded by DSS treatment is uncertain. Previous studies investigating the periodontitis-IBD interconnection lacked evidence that DSS treatment alone did not affect periodontal tissues, the oral microbiome, and oral immune cells[200,285,296,297,313-316]. This major limitation highlights the need for future research to focus on developing more accurate and representative animal models that better mimic the complex interactions between periodontitis, oral and gut bacteria, and IBD in humans. This may involve the use of alternative mouse models exhibiting increased susceptibility to both periodontitis and IBD without the need for chemical induction.

**Conclusions**

As our understanding of periodontitis and IBD pathogenesis advances, both disorders represent a complex interplay of genetic, environmental, microbial, and immunological factors. Central to this conversation is the "multi-hit" model, suggesting that a sequence of events - dysbiosis of oral microbiota, a disrupted intestinal barrier function, and an aberrant immune response - are crucial for the development of periodontitis-associated IBD. The oral cavity may serve as a potential reservoir for intestinal pathogens. Oral bacteria and their byproducts can reach the gut through ingestion or systemic circulation. Factors such as genetic disorders, diet, aging, stress, and the use of tobacco, alcohol, or medications may disrupt the intestinal barrier function, allowing the oral bacteria to colonize the gut. Once colonized, these oral bacteria can elicit excessive immune response promoting intestinal inflammation. While current studies provide valuable insights, there is a need for longitudinal studies to provide an in-depth understanding of the periodontitis-IBD interconnection. Despite the valuable insights gained from DSS-colitis animal models, their limitations must be carefully considered when exploring the interconnection between periodontitis and IBD. Lastly, collaboration between dentists, gastroenterologists, immunologists, and infectious disease experts/microbiologists, alongside other healthcare professionals, is necessary to provide holistic oral-systemic healthcare. Optimal dental care could reduce the supply of pathogenic oral bacteria to the gut, offering innovative methods to reduce the risk and severity of IBD.

**Conflict of Interest**

The authors declare that the research was conducted in the absence of any commercial or financial relationships that could be construed as a potential conflict of interest.

**Funding Sources:**

This work was supported by the National Institutes of Health grants R03DE029258 and R56DK131277 to V.T.M.; Univ. of Maryland School of Dentistry start-up funds and INSPIRE grants to V.T.M.; Merit Review Award BX004890 from the United States (U.S.) Department of Veterans Affairs Biomedical Laboratory Research and Development Program to J-P. Raufman (The contents do not represent the views of the U.S. Department of Veterans Affairs or the United States Government); and The Bloomberg~Kimmel Institute for Immunotherapy, Johns Hopkins School of Medicine (CLS).

**Author Contributions**

All authors have made substantial contributions to the following: (1) the conception and design of the study, or acquisition of data, or analysis and interpretation of data, (2) drafting the article or revising it critically for important intellectual content, and (3) final approval of the version to be submitted.

**Figure 1.** Schematic of *IRF8* gene structure with previously reported IBD single nucleotide polymorphisms (SNPs), genotypic changes, and P values.

**Figure 2. One hit, not enough wit – A multi-hit model: (red circle)** The onset of periodontitis leads to expansion of pathogenic oral bacteria. These bacteria can translocate to the intestine, either via ingestion or through the circulatory system. Normally, ingested oral bacteria are inactivated as they pass through the stomach. **(green circle)** Several factors, such as genetic disorders, medications, diet, smoking, alcohol, aging, and stress may disrupt intestinal barrier function and shift the gut microbiota long-term. This disruption can reduce colonization resistance, facilitating the establishment of oral bacteria in the gut. **(blue circle)** Once established in the gut, oral bacteria activate both innate and adaptive immune responses leading to intestinal inflammation. Oral dysbiosis or disrupted gut barrier alone are insufficient to predispose the host to IBD. Both a sufficient load of oral pathogens and disrupted barrier function are necessary for oral pathogens to successfully colonize the gut, which then activates immune responses leading to intestinal inflammation.

**Figure 3. Oral pathogen-mediated immune responses that drive gut inflammation in IBD.** Periodontitis onset triggers expansion of pathogenic oral bacteria. Constant saliva swallowing enables these bacteria to translocate to the gut. A compromised intestinal barrier function, marked by diminished mucus and epithelial barrier integrity, facilitates the penetration of oral bacteria into the sub-epithelial regions. Neutrophils, the first responders, attempt to phagocytose the ingested bacteria and release antimicrobial substances. Additionally, the antigen-presenting cells (APCs) in the gut recognize the microbial invaders through pattern recognition receptors like Toll-like receptors (TLRs). Once activated, these cells release a cocktail of pro-inflammatory molecules: interleukin (IL)-6, IL-12, IL-23, IL-1β, tumor necrosis factor α (TNF-α), and chemokines. Consequently, these molecules guide the differentiation of naive T helper (Th0) cells into Th1, Th2, Th17, and Treg cells. Furthermore, B cells become activated by recognizing bacterial antigens and differentiate into plasma cells, which produce antibodies specific to the oral bacterial antigens. These antibodies neutralize or opsonize bacteria and can form immune complexes, intensifying inflammation. The concerted actions of innate and adaptive immune cells lead to initiation or exacerbation of the inflammatory process in IBD, highlighting the far-reaching effects of oral bacterial dysbiosis. Figure created with BioRender.com.

| Table 1: Studies Demonstrating Association Between Periodontitis and IBD ||||||
| --- | --- | --- | --- | --- |
| **Study (reference number)** | **IBD (n)** | **Non-IBD (n)** | **Periodontitis-IBD Associations** | **95% CI** |
| Grossner-Schreiber et al. 2006 [179] | IBD (62) | (59) | IBD (RR = 2.47) CAL≥ 5mm | 1.02–5.99 |
| Habashneh et al. 2012 [323] | CD (59) UC (101) | (100) | CD (OR = 4.9) UC (OR = 7.0) | 1.8–13.2 2.8–17.5 |
| Vavricka et al. 2013 [139] | CD (69) UC (44) | (113) | CD (OR = 3.91) UC (OR = 3.94) | 1.78–8.57 1.64–9.46 |
| Chi et al. 2018 [181] | CD (6657) | (26,628) | (HR = 1.36) | 1.25–1.48 |
| Yu et al. 2018 [324] | CD (7) UC (20) | (108) | IBD (HR = 1.82) CD (HR = 3.95) UC (HR = 1.39) | 1.09–3.03 1.59–9.82 0.69–2.46 |
| Zhang et al. 2020 [325] | CD (265) UC (124) | (265) | CD (OR = 4.46) UC (OR = 4.66) | 2.50–7.95 2.49–8.71 |
| Bertl et al. 2022. [306] | IBD (1108) | (3429) | CD (OR = 1.74) UC (OR = 2.57) | 1.36–2.24 1.99–3.32 |
| Wang et al. 2023 [305] | IBD (12,882) | (21,770) | IBD (OR = 1.06) | 1.01–1.12 |
| Baima et al. 2023 [37] | CD (117) UC (60) | (180) | IBD (OR = 4.48) CD (OR = 4.01) UC (OR = 4.43) | 2.30–8.73 1.92–8.38 1.66–11.81 |

IBD, inflammatory bowel disease; CD, Crohn's disease; UC, ulcerative colitis; OR, odds ratio; HR, hazard ratio; RR, relative risk; CI, confidence interval; CAL, clinical attachment loss.

| | Study (reference number) | Study Design | Subjects (n) | Samples | Methods | Key Findings |
|---|---|---|---|---|---|---|
| ***Fusobacterium spp.*** *(F. nucleatum)* | Strauss et al. 2011 [267] | Case-control | UC (4), CD (17), intermediate colitis (1), Healthy controls (32), IBS (2) | Colon biopsies | Isolation, culture, 16S rRNA sequencing | *Fusobacterium* spp. abundant in patients with GI disease (64%) versus healthy controls (26%). Highly invasive strains of *F. nucleatum* enriched in IBD patients. |
| | Castellarin et al. 2012 [326] | Cross-sectional | CRC (11) | CRC biopsies and adjacent normal tissues | Isolation, culture, RNA-seq, qPCR | Higher abundance of *Fusobacterium* in colorectal tumor specimens. |
| | Dejea et al. 2014 [327] | Cross-sectional | USA CRC/Polyp (34), Healthy controls (62), MAL CRC/Polyp (22) | Intestinal biopsies from CRC and healthy controls | 16S rRNA sequencing | *Fusobacterium* predominant in non-biofilm CRC specimens and undetected in normal tissues. |
| | Gevers et al. 2014 [223] | Case-control | Pediatric CD (447), healthy controls (221) | Ileal and rectal biopsies, stool, serum | 16S rRNA sequencing | *Fusobacteriaceae* enriched in treatment naïve CD patients and associated with disease progression. Oral species better represented in biopsies vs stool. Antibiotic use amplified the microbial dysbiosis associated with CD. |
| | Drewes et al. 2017 [328] | Case-control | MAL1 CRC (21), Polyp (1), Healthy controls (34); MAL2 CRC (23), Healthy controls (23); USA CRC (35), Polyp (4), | Stool, Intestinal biopsies from CRC, Polyp, and healthy tissues | 16S rRNA sequencing | Oral pathogens, including *F. nucleatum, P. micra,* and *P. stomatis* highly enriched in CRC tissues. |

| | | | Healthy controls (60) | | | |
|---|---|---|---|---|---|---|
| | Pascal et al. 2017 [329] | Case-control | UC (74), CD (87), healthy controls (111) | Stool | 16S rRNA sequencing | *Fusobacterium* enriched in CD. |
| | Flemer et al. 2018 [330] | Case-control | CRC (99), colorectal polyps (32), Healthy controls (103) | Oral swab, colonic biopsies, stool | 16S rRNA sequencing | *F. nucleatum* enriched in CRC, but also noted in healthy controls. |
| | Schirmer et al. 2018 [331] | Observational | Pediatric UC (405) | Rectal biopsies, stool | 16S rRNA sequencing | *Fusobacterium* enriched and associated with disease progression, and twice higher in rectal biopsies vs stool. |
| | Dinakaran et al. 2018 [332] | Cross-sectional | Disease tissue from UC (13), CD (13), and adjacent healthy tissue (13) | Colonic biopsies | 16S rRNA sequencing | *Fusobacterium* enriched in diseased colonic biopsies vs adjacent healthy tissue. |
| | Komiya et al. 2019 [218] | Cross-sectional | CRC (14) | Saliva, colonic biopsies | Isolation, culture, PCR, 16S rRNA sequencing | Identical *F. nucleatum* strains found in saliva and colon tumors. |
| | Vieira-Silva et al. 2019 [333] | Case-control | PSC (64), UC (13), CD (29), Healthy (1,120) | Stool | 16S rRNA sequencing | *Fusobacterium* abundance associated with severity of intestinal inflammation. |
| | Liu et al. 2020 [285] | Case-control | UC (20), CD (71), healthy controls (43) | Stool | 16S rRNA sequencing | *F. nucleatum* enriched and associated with disease activity in UC and CD patients. *F. nucleatum* infection in DSS-treated mice exacerbated colitis by promoting expression of IL-1β, TNF-α, IFN-γ, IL-6, and IL-17, and fostering CD4+T cell differentiation into |

| | | | | | | Th1 and Th17 cells. |
|---|---|---|---|---|---|---|
| *Porphyromonas spp.*<br><br>*(P. gingivalis)* | Flemer et al. 2017 [334] | Case-control | CRC (59), colon polyps (21), healthy controls (56) | Stool, CRC biopsies and adjacent normal tissues | 16S rRNA sequencing | Microbiota differed in CRC vs controls. Alterations were noted throughout the colon and not restricted to cancer. |
| | Ahn et al. 2013 [335] | Case-control | CRC (47), healthy controls (94) | Stool | 16S rRNA sequencing | *Fusobacterium* and *Porphyromonas* were enriched in CRC, whereas *Clostridia* were decreased. |
| | Yang et al. 2019 [336] | Case-control | CRC (50), healthy controls (50) | Stool | 16S rRNA sequencing, gas chromatography-mass spectrometry | CRC had less microbial diversity and enriched *Fusobacteria, Parvimonas,* and *Porphyromonas*; 17 metabolites were specifically perturbed in CRC. |
| | Wang et al. 2021 [337] | Cross-sectional | Healthy (22), CRC (23), colorectal adenoma (32) | Stool, CRC, and normal tissue biopsies | 16S rRNA sequencing, qPCR | *P. gingivalis* enriched in stool and biopsies of CRC and associated with poor prognosis. In $Apc^{Min/+}$ mice, *P. gingivalis* promoted CRC by activating NLRP3 inflammasome. |
| | Lee et al. 2022 [338] | Case-control | CD (11), control (8) | Stool | 16S rRNA sequencing | *Porphyromonadaceae* enriched in CD. *P. gingivalis* infection in DSS-treated mice exacerbated colitis by promoting expression of TNF-α and IL-6. |
| *Aggregatibacter spp.* | Dinakaran et al. 2018 [332] | Cross-sectional | Disease tissue from UC (13), CD (13), and adjacent healthy tissue (13) | Colon biopsies | 16S rRNA sequencing | *Aggregatibacter, Porphyromonas, Prevotella, Staphylococcus, Streptococcus,* enriched in diseased colon |

| | | | | | | biopsies vs adjacent healthy tissue. |
|---|---|---|---|---|---|---|
| *Prevotella spp.* | Atarashi et al. 2017 [229] | Case-control | UC (51), CD (7), PSC (27), GERD (18), alcoholic (16), healthy controls (150) | Saliva, Stool | Isolation, culture, 16S rRNA sequencing, metagenomic analysis | *Prevotella, Streptococcus, Neisseria, Rothia,* and *Gemella* enriched in UC, PSC, GERD and alcoholic feces. |
| | Dinakaran et al. 2018 [332] | Cross-sectional | Disease tissue from UC (13), CD (13), and adjacent healthy tissue (13) | Colon biopsies | 16S rRNA sequencing | *Prevotella* enriched in diseased colon vs adjacent healthy tissue. |
| | Prasoodanan et al. 2021 [339] | Population-based | IBD (189) and healthy controls (200). | Stool | Whole metagenome sequencing | Higher abundance and diversity of *P. copri* in Indian and non-western versus western populations. |
| *Veillonellaceae spp.* (*V. parvula*) | Gevers et al. 2014 [223] | Case-control | Pediatric CD (447), healthy controls (221) | Ileal and rectal biopsies, stool, serum | 16S rRNA sequencing | *Veillonellaceae* enriched in CD and associated with disease progression. Oral species better represented in biopsies vs stool. |
| | Schirmer et al. 2018 [331] | Observational | Pediatric UC (405) | Rectal biopsies, stool | 16S rRNA sequencing | *V. parvula* enriched and associated with disease progression, and higher in rectal biopsies vs stool. |
| | Vieira-Silva et al. 2019 [333] | Case-control | PSC (64), UC (13), CD (29), Healthy (1,120) | Stool | 16S rRNA sequencing | *Veillonella* abundance associated with severity of intestinal inflammation. |
| *Klebsiella* spp. (*K. pneumoniae*) | Atarashi et al. 2017 [229] | Case-control | UC (51), CD (7), PSC (27), GERD (18), alcoholic (16), healthy controls (150) | Saliva, Stool | Isolation, culture, 16S rRNA sequencing, metagenomic analysis | *Klebsiella* enriched in CD. Antibiotic-resistant *Klebsiella* tend to colonize when gut microbiota is dysbiotic and elicit severe inflammation in a genetically |

| | | | | | | susceptible host by promoting Th1 cells. |
|---|---|---|---|---|---|---|
| | Lloyd-Price et al. 2019 [340] | Case-control | UC (38), CD (67), healthy controls (27) | Colon biopsies, stool, blood | 16S rRNA sequencing, metagenomic, metatranscriptomic, proteomic, metabolomic analysis | *K. pneumoniae* and *H. parainfluenzae* levels during dysbiosis associated with acylcarnitines and bile acid levels. |
| | Imai et al. 2021 [308] | Case-control | UC (42), CD (18), healthy controls (45) | Saliva and stool | 16S rRNA sequencing | *Enterobacteriaceae* enriched in CD patients with periodontitis. |
| *Streptococcus spp.* | Pascal et al. 2017 [329] | Case-control | UC (74), CD (87), healthy controls (111) | Stool | 16S rRNA sequencing | Higher abundance of *Streptococcus* noted in CD with post-operative recurrence compared to those who remained in remission. |
| | Atarashi et al. 2017 [229] | Case-control | UC (51), CD (7), PSC (27), GERD (18), alcoholic (16), healthy controls (150) | Saliva, Stool | Isolation, culture, 16S rRNA sequencing, metagenomic analysis | *Streptococcus,* enriched in feces of UC, PSC, GERD, and alcoholic patients. |
| | Vieira-Silva et al. 2019 [333] | Case-control | PSC (64), UC (13), CD (29), Healthy (1,120) | Stool | 16S rRNA sequencing | *Streptococcus* abundance associated with severity of intestinal inflammation. |
| *Campylobacter spp.* (*C. concisus*) | Man et al. 2010 [266] | Case-control | Pediatric CD (54), non-IBD (27), and healthy controls (33). | Stool | PCR | Higher prevalence of *C. concisus* noted in pediatric CD versus non-IBD and healthy controls. |
| | Kirk et al. 2016 [341] | Case-control | UC (16), CD (9), IPAA (27), and healthy controls (26) | Ileum and colon biopsies | Isolation, culture, and PCR | Higher abundance of *C. concisus* noted in IBD versus control. |

IBD, inflammatory bowel disease; CD, Crohn's disease; UC, ulcerative colitis; IBD, irritable bowel syndrome; CRC, colorectal cancer; PSC, primary sclerosing cholangitis; GERD, gastroesophageal reflux disease; GI, gastrointestinal; MAL1, Malaysian cohort 1; MAL2, Malaysian cohort 2; DSS, dextran sodium sulfate; IPAA, ileal pouch–anal anastomosis.

**Figure 1**

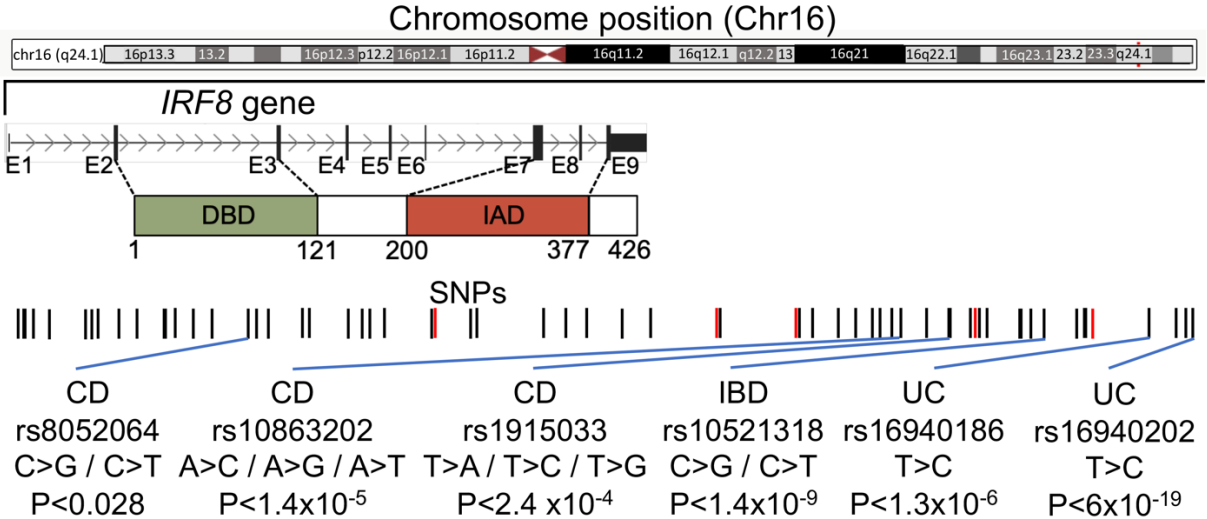

**Figure 1.** Schematic of *IRF8* gene structure with previously reported IBD single nucleotide polymorphisms (SNPs), genotypic changes, and P values.

**Figure 2**

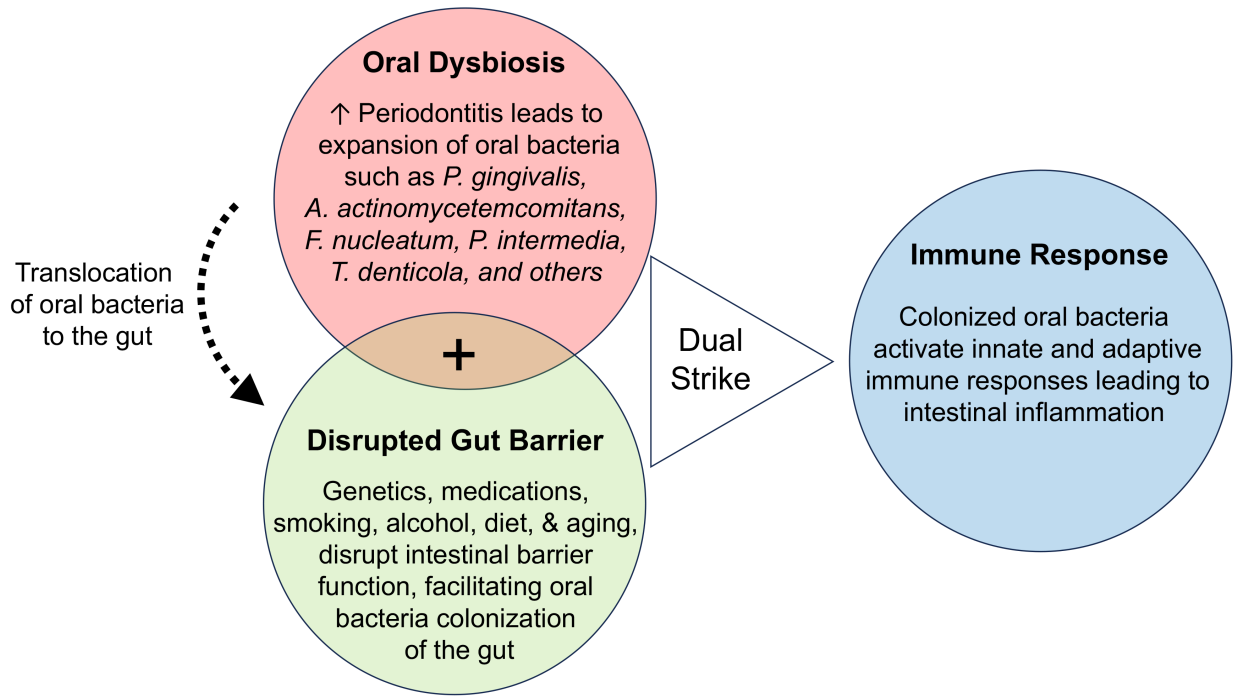

**Figure 2. One hit, not enough wit – A multi-hit model: (red circle)** The onset of periodontitis leads to expansion of pathogenic oral bacteria. These bacteria can translocate to the intestine, either via ingestion or through the circulatory system. Normally, ingested oral bacteria are inactivated as they pass through the stomach. **(green circle)** Several factors, such as genetic disorders, medications, diet, smoking, alcohol, aging, and stress may disrupt intestinal barrier function and shift the gut microbiota long-term. This disruption can reduce colonization resistance, facilitating the establishment of oral bacteria in the gut. **(blue circle)** Once established in the gut, oral bacteria activate both innate and adaptive immune responses leading to intestinal inflammation. Oral dysbiosis or disrupted gut barrier alone are insufficient to predispose the host to IBD. Both a sufficient load of oral pathogens and disrupted barrier function are necessary for oral pathogens to successfully colonize the gut, which then activates immune responses leading to intestinal inflammation.

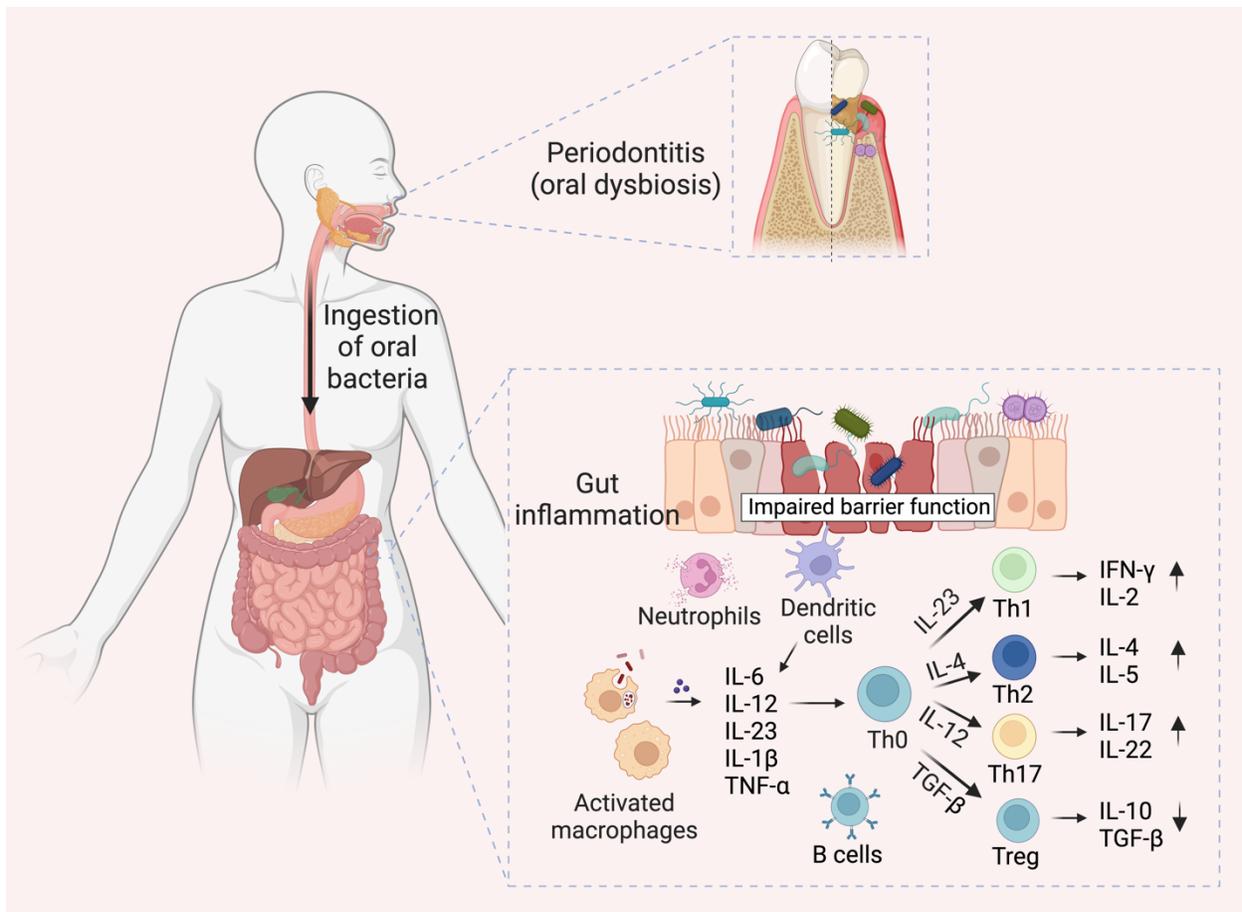

**Figure 3. Oral pathogen-mediated immune responses that drive gut inflammation in IBD.** Periodontitis onset triggers expansion of pathogenic oral bacteria. Constant saliva swallowing enables these bacteria to translocate to the gut. A compromised intestinal barrier function, marked by diminished mucus and epithelial barrier integrity, facilitates the penetration of oral bacteria into the sub-epithelial regions. Neutrophils, the first responders, attempt to phagocytose the ingested bacteria and release antimicrobial substances. Additionally, the antigen-presenting cells (APCs) in the gut recognize the microbial invaders through pattern recognition receptors like Toll-like receptors (TLRs). Once activated, these cells release a cocktail of pro-inflammatory molecules: interleukin (IL)-6, IL-12, IL-23, IL-1β, tumor necrosis factor α (TNF-α), and chemokines. Consequently, these molecules guide the differentiation of naive T helper (Th0) cells into Th1, Th2, Th17, and Treg cells. Furthermore, B cells become activated by recognizing bacterial antigens and differentiate into plasma cells, which produce antibodies specific to the oral bacterial antigens. These antibodies neutralize or opsonize bacteria and can form immune complexes, intensifying inflammation. The concerted actions of innate and adaptive immune cells lead to the initiation or exacerbation of inflammatory process in IBD, highlighting the far-reaching effects of oral bacterial dysbiosis. Figure created with BioRender.com.